\documentclass[conference]{IEEEtran}
\IEEEoverridecommandlockouts

\usepackage{cite}
\usepackage{amsmath,amssymb,amsfonts}
\usepackage{algorithmic}
\usepackage{graphicx}
\usepackage{textcomp}
\usepackage{float}
\usepackage{xcolor}
\usepackage{hyperref}
\usepackage{listings}
\usepackage{cleveref}
\usepackage{pifont}
\usepackage{xspace}
\usepackage{tabularx} 
\usepackage[numbers]{natbib}
\usepackage{enumitem}
\usepackage{booktabs}
\usepackage{pifont}
\usepackage{arydshln}
\usepackage{makecell}
\usepackage{tikz}
\usepackage{subcaption}
\usepackage{soul}

\DeclareRobustCommand{\cb}[6]{%
  \tikz[baseline={(r.base)}]{
    \node[inner sep=0pt, outer sep=0pt] (r) {};%
    \begin{scope}[yshift=-2pt] 
      \draw[
        fill={rgb,1:red,#1; green,#2; blue,#3},
        draw={rgb,1:red,#4; green,#5; blue,#6},
        line width=0.4pt
      ] (0,0) rectangle (8pt,8pt);
    \end{scope}
  }%
}

\usepackage{balance}

\newcommand{\ie}{i.e.,\xspace}
\newcommand{\eg}{e.g.,\xspace}

\mathchardef\UrlBreakPenalty=1000
\mathchardef\UrlBigBreakPenalty=1000

\newcolumntype{L}[1]{>{\raggedright\arraybackslash}p{#1}}
\newcolumntype{C}[1]{>{\centering\arraybackslash}p{#1}}
\newcolumntype{R}[1]{>{\raggedleft\arraybackslash}p{#1}}

\def\BibTeX{{\rm B\kern-.05em{\sc i\kern-.025em b}\kern-.08em
    T\kern-.1667em\lower.7ex\hbox{E}\kern-.125emX}}
\begin{document}

\title{Performance Antipatterns: \\Angel or Devil for Power Consumption?}

 \author{
    \IEEEauthorblockN{Alessandro Aneggi}
    \IEEEauthorblockA{\textit{Free University of Bozen-Bolzano}\\
    Bolzano, Italy \\
    aaneggi@unibz.it}

\and
    \IEEEauthorblockN{Vincenzo Stoico}
    \IEEEauthorblockA{Vrije Universiteit Amsterdam \\
    Amsterdam, Nederland \\
    v.stoico@vu.nl}
    
    \and
    \IEEEauthorblockN{ Andrea Janes}
    \IEEEauthorblockA{\textit{Free University of Bozen-Bolzano}\\
    Bolzano, Italy \\
    ajanes@unibz.it}

}

\maketitle


\begin{abstract}
Performance antipatterns are known to degrade the responsiveness of microservice-based systems, but their impact on energy consumption remains largely unexplored. This paper empirically investigates whether widely studied performance antipatterns defined by Smith and Williams also negatively influence power usage. We implement ten antipatterns as isolated microservices and evaluate them under controlled load conditions, collecting synchronized measurements of performance, CPU and DRAM power consumption, and resource utilization across 30 repeated runs per antipattern. The results show that while all antipatterns degrade performance as expected, only a subset exhibit a statistically significant relationship between response time and increased power consumption. Specifically, several antipatterns reach CPU saturation, capping power draw regardless of rising response time, whereas others (\eg Unnecessary Processing, The Ramp) demonstrate energy-performance coupling indicative of inefficiency. Our results show that, while all injected performance antipatterns increase response time as expected, only a subset also behaves as clear energy antipatterns, with several cases reaching a nearly constant CPU power level where additional slowdowns mainly translate into longer execution time rather than higher instantaneous power consumption. The study provides a systematic foundation for identifying performance antipatterns that also behave as energy antipatterns and offers actionable insights for designing more energy-efficient microservices architectures. 
\end{abstract}


\maketitle

\section{Introduction}
Software systems often experience recurring conditions that negatively impact their performance.
Recognizing, avoiding, and resolving these issues is vital for maintaining high performance and ensuring software sustainability over time and across contexts.
Since their introduction in 2002 \cite{Smith2002}, software performance antipatterns have been the subject of extensive study across different domains of software engineering.
Current research emphasizes early detection of antipatterns during the software engineering process, especially at architectural level, to reduce future issues and avoid costly refactoring in production \cite{cortellessa2023many, avritzer2025architecture, matar2023approach, Pinciroli2023, Avritzer2021, van2024slow}.

Despite the recognized importance of antipatterns for performance optimization, \textit{research on their effect on software energy efficiency remains limited}~\cite{araujo2024energy}.
Architectural decisions, however, can substantially influence energy efficiency, as demonstrated by~\citet{zhao2025does}.
\citet{zhao2025does} reveal a complex interplay between energy consumption and performance in microservice architectures.  
Specifically, finer service granularity tends to increase both energy usage and response time compared to coarser designs.  
Moreover, higher workloads, represented as increased request rates, further amplify both energy consumption and performance costs.  
The study underscores that the relationship between energy efficiency and performance, shaped by service granularity and system scale, must be carefully evaluated during the design phase.

At the current stage, \textit{it remains unclear whether performance antipatterns are also harmful to software energy efficiency} \cite{Morales2016}.
Establishing such evidence would be valuable for software architects, enabling them to assess performance-energy tradeoffs during the design phase.
Performance indicators may be independent of energy use, with the former being favored in contexts where performance is assessed primarily through output quality or throughput rather than energy efficiency.
This study examines the relationship between performance and energy efficiency, focusing on response time and power consumption.
The analysis could highlight scenarios where prolonged response time leads to energy waste, while extended response time does not always correspond to increased power consumption, such as in case of frequent low intensity workloads.
Analyzing power consumption helps isolate performance trade-offs by removing the influence of time from the energy analysis \cite{aragon2025does}. 
Since energy represents the total power consumed over a given duration, and response time defines that duration, focusing on power directly avoids conflating energy efficiency with timing effects, helping in identifying potential waste.

Our \textbf{goal} is to identify the cases in which performance antipatterns lead to energy waste.
We design and conduct a controlled experiment based on the ten performance antipatterns described by ~\citet{smith2003more}.
These antipatterns are manually implemented as a container and profiled under a fixed workload and observation period of 20 minutes.
This experimental setup enables us to examine the relationship between performance and energy metrics while reducing potential confounding factors that could obscure causal effects.
We measure response time, power consumption, and resource utilization.
The resource utilization data allow us to verify whether the observed behavior aligns with the antipattern definitions, as the original descriptions provide no implementation guidelines and are intended to remain implementation-agnostic.

Our \textbf{results} show that the occurrence of performance antipatterns does not necessarily lead to energy waste.
The Ramp, God Class, and Traffic Jam antipatterns increase both response time and power consumption, while also saturating CPU usage.
In these scenarios, the prolonged response time causes energy waste as a consequence of performance inefficiencies.
For the other antipatterns, the increase in power consumption does not significantly affect response time. 
In such cases, response time remains stable, while power and CPU resources are still unnecessarily drained.
The Unnecessary Work antipattern exemplifies this behavior.

The \textbf{main contributions} of this study are as follows:
(i) an empirical analysis of the relationship between response time and power consumption in the presence of performance antipatterns;
(ii) an in-depth discussion of the findings, emphasizing the trade-offs between energy efficiency and software performance; and
(iii) a replication package containing the raw data, analysis scripts, and source code to facilitate full reproducibility of the study (see \cite{replication_package}).

Our findings reveal that performance antipatterns are not inherently detrimental to energy efficiency.  
Consequently, software architects should carefully assess the trade-offs between performance and energy consumption during design-time.
For researchers, the study underscores the importance of treating response time and energy efficiency as independent dimensions of analysis.  
Finally, our work bridges the two communities, illustrating how insights from software performance evaluation can enrich software architecture design.


\section{Related Work}
\label{sec:related}

The term \textit{Antipattern} was coined by Andrew Koenig in 1995 to describe solutions to recurring problems that superficially look like good solutions but later reveal to have disadvantages, which make it a bad solution \cite{Rising1998,Long2001}.

Software \textit{performance antipatterns} are commonly design solutions that are actually counterproductive to system performance. Starting from 2000, \citet{Smith2002} identified several families of performance antipatterns and, in 2002, proposed a catalog of recurring design problems that systematically lead to performance issues. Since then, additional antipatterns have been identified and classified. Since these antipatterns form the starting point of this paper, we report the antipatterns defined there in Tab. \ref{tab:antipatterns}.

\begin{table}[htbp]
\setlength{\tabcolsep}{2pt}
\centering
\caption{Performance Antipatterns discussed in \cite{Smith2002}}
\label{tab:antipatterns}
\begin{tabular}{@{}L{2.3cm}L{6.4cm}@{}}
\toprule
\textbf{Antipattern} & \textbf{Problem} \\
\midrule

Unbalanced Processing & Processing does not use available processors; one slow stage limits throughput. \\\hdashline
Unnecessary Processing & Work is executed although it is not required at that moment or not required at all. \\\hdashline
The Ramp & Processing time grows with system usage or data size. \\\hdashline
Sisyphus DB Retrieval & Full queries compute more data than needed. \\\hdashline
More is Less & Too many processes cause thrashing instead of progress. \\\hdashline
God Class & One class concentrates too much work or data. \\\hdashline
Excessive Dynamic Allocation & Too many short-lived objects are created and destroyed. \\\hdashline
Circuitous Treasure Hunt & Data must be collected from many locations with high lookup cost. \\\hdashline
One-Lane Bridge & Only one process can continue; others must wait. \\\hdashline
Traffic Jam & A temporary issue causes long-lasting request backlogs. \\

\bottomrule
\end{tabular}
\end{table}

Performance antipatterns have been studied from different perspectives.  \citet{Smaalders2006} describes them as pathologies that can manifest across multiple levels of a software system, in code produced by both researchers and practitioners. 
%
The impact of antipatterns on performance have also been evaluated using models (i.e., Queuing Networks) at design-time \cite{Pinciroli2023, Cortellessa2012}.
~\citet{Avritzer2021} introduce a computationally efficient multivariate approach for characterizing and detecting performance antipatterns, designed to support CI/CD pipelines.

\textit{Software energy consumption} can be measured using physical power meters, 
via on-chip energy counters, or by means of software-only estimators based on analytical or statistical models~\cite{Kruglov2023}. The introduction of RAPL in modern CPU architectures has significantly increased the possibility of collecting fine-grained energy data in software and attributing it to specific workloads or software components~\cite{Hahnel2012}. Building on these capabilities, several recent works focus on software-, server-, and container-level observability of energy and power consumption~\cite{pijnacker2025,Amaral2024,Fieni2020}.

\textit{The relationship between performance and energy efficiency} in server-based systems is non-trivial. From a software engineering perspective, empirical studies confirm that design decisions impact energy and performance differently: \citet{Noureddine2025} show that design patterns often increase energy consumption despite performance intentions, while refactoring code smells can improve energy efficiency by 49\% or increase it by 70\% depending on the smell type~\cite{verdecchia2018empirical}. \citet{procaccianti2016empirical} demonstrate that best practices making CPU usage more energy-proportional can reduce consumption up to 25\%, but parallel execution improves the energy-performance trade-off compared to sequential execution.
In different contexts, researchers find discrepancies between energy and performance metrics. Code smell refactoring shows mixed results~\cite{verdecchia2018empirical}, while microservice techniques present complex trade-offs: reducing energy by up to 5.6\% can increase latency by 25.9\% in some cases, while decreasing it by 72.5\% in others~\cite{Xiao2025}. Architectural choices also affect this phenomenon, as demonstrated by microservice granularity studies~\cite{zhao2025does}.
These results suggest that performance problems and energy inefficiencies are often correlated but not perfectly aligned, motivating an explicit analysis of software performance antipatterns from the energy-consumption perspective.

\textit{Recent research establishes energy as a first-class quality attribute} in software architecture, requiring explicit consideration throughout the design process~\cite{Bashroush2017,Beik2012}. Foundational work demonstrates that energy-aware architectural layers and design principles are essential for system-level optimization. A key observation is that energy-optimal resource allocation configurations often diverge significantly from performance-optimal solutions~\cite{Xu2016}, necessitating dedicated energy-aware resource management strategies. Industrial empirical studies validate the effectiveness of architectural tactics in practice: \citet{Funke2024} demonstrate that deployment model selection and granular auto-scaling can achieve energy reductions ranging from 21-50\% under typical loads. Beyond architectural decisions, complementary optimization approaches at lower abstraction levels (including compiler-based techniques~\cite{Suresh2014} and machine learning-driven resource allocation~\cite{Agomuo2024}) extend energy efficiency improvements across implementation strategies. In microservice contexts, balancing energy efficiency with failure resilience remains a critical concern~\cite{Ponce2025}. While evidence confirms that energy-aware design is broadly applicable across diverse application domains, adoption barriers persist in industrial practice due to limited measurement transparency and the need for systematic decision-making.

\textit{Performance antipattern detection} employs multiple approaches across abstraction levels. Model-based frameworks like PADRE~\cite{Cortellessa2022} identify antipatterns at design time through UML analysis with automated refactoring; runtime statistical methods~\cite{Avritzer2021} enable CI/CD integration. 
\textit{Energy measurement} is based the two main methodologies: monitoring (for CPU, DRAM and GPU) or estimation (which add also Network and Disk) \cite{Priyavanshi2025}. 
Monitoring relies on Intel's Running Average Power Limit (RAPL) which is a hardware interface built into modern Intel processors that provides fine-grained energy consumption measurements at runtime for CPU, uncore components, and DRAM through model-specific registers\cite{Khan2018}.
Various tools are available based on RAPL with varying level of granularity, such as Scaphandre~\cite{Scaphandre2020}, which provides per-process attribution across bare metal and containerized environments; PowerJoular~\cite{noureddine-ie-2022}, which enables process-level Linux monitoring; PowerLetrics~\cite{Geerd-Dietger2025}, which  offers real-time energy footprints; and JoularJX~\cite{noureddine-ie-2022}, which targets Java-level granularity. However, \citet{Morales2016} reveal a critical phenomenon: seven of the studied object-oriented and mobile antipatterns improve energy efficiency of an Android app when removed, while two paradoxically worsen it, indicating that performance optimization and energy efficiency are not perfectly aligned. Despite mature detection tools for performance and a comprehensive measurement infrastructure for energy, no integrated approach combines both perspectives to detect energy-specific antipatterns, particularly in microservice contexts where granularity and resource allocation create novel energy-performance trade-offs.


\section{Study Design and Hypothesis Formulation}
\label{sec:hypothesis}

This work investigates whether the well-known performance antipatterns defined by \citet{smith2003more}, beyond degrading system performance, may also serve as indicators of increased energy consumption in microservices.

The study aims to deepen the understanding of the relationship between energy usage and performance, focusing on the impact of performance antipatterns within containers. To operationalize this objective, we quantify energy usage through direct measurements of power consumption in containerized environments. Therefore, we derive the following research question: \textbf{how does the presence of performance antipatterns influence the power consumption of microservices?}

\begin{figure}[htbp]
  \centering
  \includegraphics[width=\linewidth]{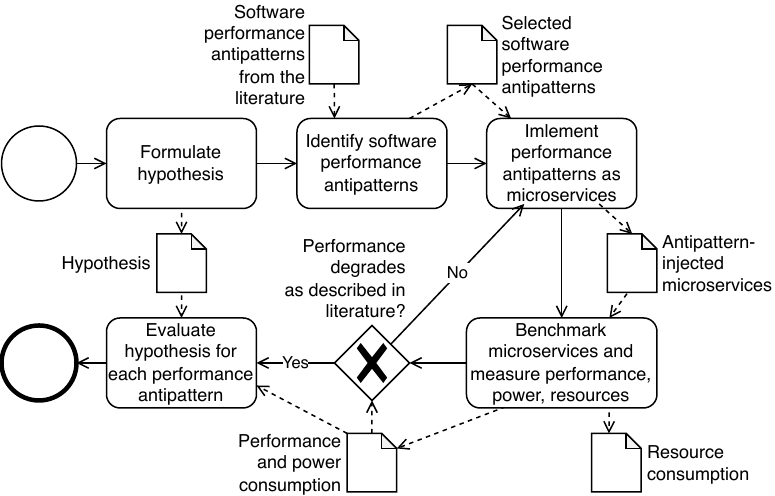}
  \caption{Research Process (in BPMN \cite{bpmn} notation).}
  \label{fig:researchprocess}
\end{figure}

Figure \ref{fig:researchprocess} depicts the research methodology followed in this study: we begin with the hypothesis formulation (see Sect. \ref{sec:hypothesis}) and the study of the related work (see Sect. \ref{sec:related}). As a next step, we identify the software antipatterns we want to evaluate and implement them as microservices (see Sect. \ref{ref:package}). We proceed load-testing each antipattern-injected microservice to extract three key metric groups: performance, power, and resource consumption metrics. We then evaluate analytically if the performance of the antipattern-injected microservices degrade as described in the literature, \ie if they constitute a valid example for the studied antipattern. If not, the implementation is revised, if yes, we use the collected data to evaluate the hypothesis (see Sect. \ref{sec:analysis}).

To investigate whether microservices that experience performance degradation also exhibit higher power consumption, we formulate the following hypotheses. The \textbf{null hypothesis (H$_0$)} states that no significant relationship exists between system performance and power usage; in other words, variations in response time do not correspond to meaningful changes in power consumption. The \textbf{alternative hypothesis (H$_1$)} posits a significant relationship, whereby higher response times are associated with increased or decreased power demand.

To test these hypotheses, we collect two synchronized data streams: performance metrics (response time and throughput under varying load) and instantaneous power consumption of the system under test. The objective is to determine whether periods of degraded performance coincide with elevated power usage, indicating that performance problems are accompanied by energy inefficiency. To improve internal validity, we additionally record resource-utilization metrics (CPU, memory, network, and disk usage). This third stream allows us to verify that the microservice exhibits the resource-usage patterns associated with the injected antipatterns, ensuring that each implementation faithfully represents its intended class.

\section{Performance Antipatterns Implementation}
\label{ref:package}

This section presents the performance antipatterns selected for evaluation. We base our implementations on the antipatterns described by Smith \cite{smith2003more}, whose catalog is widely recognized in the performance engineering community. We report the antipattern and the caused problem in Tab. \ref{tab:antipatterns}.

The implementations follow these principles: execution restricted to a single CPU core, implementation as a single endpoint within a Python Flask \cite{flask} application, faithful reproduction of the antipattern's described behavior, and deployment within a single container. 
A SQLite port of the Northwind database \cite{northwind-sqlite3} was employed only for the Circuitous Treasure Hunt, Excessive Dynamic Allocation, God Class, and Sisyphus Retrieval experiments.
 The implementation and the rationale why we consider the implementation to be representative for the given antipattern are reported in Tab. \ref{tab:implementations_rationale}.

\begin{table*}[tb]
\setlength{\tabcolsep}{2pt}
\centering
\caption{Antipatterns, Implementation, and Rationale}
\label{tab:implementations_rationale}
\begin{tabular}{@{}L{1.5cm}L{5.5cm}L{10.8cm}@{}}
\toprule
\textbf{Antipattern} & \textbf{Implementation} & \textbf{Rationale} \\
\midrule

Unbalanced Processing & Repeats redundant hashing, validation, sorting, and copying, grows an unbounded in-memory store, concatenates and slices it, and applies frequent hashing and JSON churn. & The implementation performs a computationally intensive operation directly on the request path, monopolizing the processing thread. This synchronous workload prevents the server from handling concurrent requests, forcing all arrivals to wait until the initial task completes and thereby transforming potential parallelism into strict serialization. \\\hdashline

Unnecessary Processing & Runs a CPU-heavy math loop with trigonometry, exponentiation, and prime checks for a fixed iteration count, monopolizing the CPU before returning a result. & The implementation repeatedly executes operations that are not required for correctness, including redundant hashing, re-validation, sorting, string manipulation, and deep copying. These unnecessary tasks consume processor time and memory resources, increasing response time without providing functional value. \\\hdashline

The Ramp & Appends a random item to an in-memory list, chooses a random target value, linearly searches and sorts matches, and returns the findings with the current list size. & Processing time increases proportionally with accumulated system state because each request initiates a linear-time search over an expanding data structure. Although the effect appears negligible under limited testing, it becomes increasingly detrimental as data volume grows, demonstrating a failure to maintain stable performance under scaling conditions. \\\hdashline

Sisyphus DB Retrieval & Scans the entire Orders and Customers tables in a database, constructs objects for every row, then selects the requested page from that processed list. & The implementation conducts full table scans and joins for each request, materializing entire result sets even though only a small subset is ultimately returned. Most retrieved data is immediately discarded. This behavior imposes repetitive and wasteful computation that scales with dataset size while yielding limited functional output. \\\hdashline

More is Less & Reads thread and iteration counts, spawns many CPU-bound worker threads and compares the work against a single-threaded run. & Increasing the number of threads leads to reduced throughput because the overhead associated with scheduling, synchronization, and resource contention exceeds any potential performance benefit. The system spends more time coordinating threads than performing useful work, resulting in a net decline in efficiency. \\\hdashline

God Class & A single monolithic handler parses the payload, updates request and error counters, performs expensive hashing, consults a database cache, optionally executes full processing and storage, and returns the data with summary statistics. & The implementation centralizes validation, logging, caching, storage, and hashing responsibilities within a single class. Each method invocation opens new database connections, increasing message traffic and enforcing serialized input–output operations. This concentration of responsibilities degrades throughput and violates established principles of modularity and locality. \\\hdashline

Excessive Dynamic Allocation & Processes a request in two phases: first repeatedly allocates fresh containers and strings per iteration, then repeats the work while reusing prior allocations. & The implementation repeatedly instantiates short-lived objects, including lists, dictionaries, strings, and complex nested structures. These objects are created and discarded within tight loops, imposing significant allocation and memory-management overhead while providing no corresponding functional benefit. \\\hdashline

Circuitous Treasure Hunt & Handles a request by looking up a customer and recent orders, iterating over each order to pull related details, products, and suppliers, then returning the assembled result. & The implementation issues sequential and nested queries in which each intermediate result triggers additional lookups across multiple tables. This produces an extended dependency chain that requires numerous database round trips, prevents query optimization, and forces serialized execution, thereby imposing significant performance penalties. \\\hdashline

One-Lane Bridge & Serializes all callers on a global lock, performs CPU-intensive hashing while holding it, and updates shared state. & The implementation enforces exclusive access to shared resources through locking mechanisms that surround computationally intensive work. The prolonged lock duration forces other requests to wait, transforming available concurrency into a serialized queue and significantly reducing throughput. \\\hdashline

Traffic Jam & Schedules normal or heavy CPU work in windows with lingering backlog, calibrates and runs the kernel accordingly, and replies with diagnostic state. & Periodic bursts of heavy computation create a backlog whose effects persist even after the high-load interval ends. The accumulated waiting time elevates response response time for subsequent normal requests, producing extended variability in performance that remains long after the initiating event has subsided. \\

\bottomrule
\end{tabular}
\end{table*}

\section{Experiment Design} 
\label{sec:design}

We design the experiment using a \textit{within-subject design} \cite{wohlin2024experimentation}, where each subject (in this case, each antipattern implementation) is exposed to different conditions, namely a set of concurrent users in our setting.
The concurrent users corresponds to our unique \textbf{independent variable}, which we control to a single treatment.

This value is chosen according to the CPU usage, which it is induced by the amount of concurrent users requesting the service.
To ensure that the performance antipattern can actually manifest during experimentation, the system must operate under a sufficiently high CPU load. For this reason, we select a number of concurrent users that drives CPU utilization to at least 25 percent and does not lead to any execution errors. This threshold provides the necessary level of system saturation for revealing behaviors that typically arise under medium or high load, such as slowdowns.
For all antipatterns, 50 concurrent users were selected, except Unbalanced Processing (10 users), Unnecessary Processing (30 users), and Traffic Jam (30 users); this choice depends on the available hardware configuration, which is described in Sect. \ref{sec:execution}. 

The considered \textbf{dependent variables} are power consumption and performance.
Power consumption refers to the power requested by the microservice from the CPU and memory, measured in Watts (W).  
Performance is assessed through response time in milliseconds (ms). CPU usage, memory consumption, disk activity, and network utilization are collected not as performance metrics, but to verify the correctness of the antipattern implementation and to confirm its alignment with the descriptions found in the literature.
The tools used in the study, together with their roles and deployment locations, are listed in Tab.~\ref{tab:tools}. The collected metrics are presented in Tab.~\ref{tab:metrics}, along with their measurement unit and scope, an indication of whether each metric is directly collected or derived, and whether it is a point or cumulative metric.

Each trial, which includes a specific antipattern and number of concurrent users, is repeated 30 times to ensure measurement consistency and mitigate random variations from the execution environment, e.g., possible (unintentionally not closed) background operating system processes. 

Each experiment consists of a 20-minute test phase. Between experiments, we reset the environment by wiping all Docker containers, restarting the Docker service, waiting 10 seconds, deploying the solution, and waiting an additional 10 seconds. After each test, we include a 30-second cool-down period. The workload is generated with a spawn rate of 10 users per second. Considering these timings, the total duration of the full experimental campaign can be expressed as \( (20\,\text{min} + \text{setup and cool-down}) \times 10\,\text{antipatterns} \times 30\,\text{repetitions} \), which results in approximately 110 hours (4.58 days) of execution time.

\begin{table}[htbp]
\setlength{\tabcolsep}{2pt}
\centering
\caption{Components, Role, and Deployment Location.}
\label{tab:tools}
\begin{tabular}{@{}L{2cm}L{6.3cm}l@{}}
\toprule
\textbf{Tool} & \textbf{Role} & \makebox[6pt][r]{\textbf{Installation location\textsuperscript{g}}} \\
\midrule

PPTAM \cite{Avritzer2019}\textsuperscript{a} & Orchestrates and executes test scenarios based on deployment scripts and test configurations. & D \\\hdashline
PPTAM Agent\textsuperscript{b}  & Collects and aggregates data from other tools so that PPTAM can collect it. & T \\\hdashline
locust\textsuperscript{c} & Generates synthetic API traffic to load the system according to a predefined test script and collects the resulting response times per endpoint. & D \\\hdashline
cAdvisor\textsuperscript{d} & Provides access to container-level resource usage and performance metrics. & T \\\hdashline
PowerJoular \cite{Noureddine2022} & Provides access to per-process power consumption. & T \\\hdashline
perf\textsuperscript{e} & Provides access to hardware performance counters with configuration for DRAM energy estimation. & T \\\hdashline
Docker\textsuperscript{f} & Hosts the containerized software under test. & T\\

\bottomrule
\multicolumn{3}{l}{\textsuperscript{a}~\url{https://github.com/pptam/pptam-tool}, } \\
\multicolumn{3}{l}{\textsuperscript{b}~\url{https://github.com/pptam/pptam-agent}} \\
\multicolumn{3}{l}{\textsuperscript{c}~\url{https://locust.io/}} \\
\multicolumn{3}{l}{\textsuperscript{d}~\url{https://github.com/google/cadvisor}} \\
\multicolumn{3}{l}{\textsuperscript{e}~\url{https://web.eece.maine.edu/~vweaver/projects/perf_events/}} \\
\multicolumn{3}{l}{\textsuperscript{f}~\url{https://www.docker.com/}} \\
\multicolumn{3}{l}{\textsuperscript{g}~\ul{D}river, \ul{T}estbed}
\end{tabular}
\end{table}

\begin{table}[t]
\setlength{\tabcolsep}{.7pt}
\centering
\caption{Measurements, units, and their classification}
\label{tab:metrics}
\begin{tabular}{@{}l@{\hspace{2pt}}L{3.22cm}L{2cm}ccccc cc@{}}
\toprule
\textbf{Tool} & \textbf{Metric} & \makecell[l]{\textbf{Unit}} & \rotatebox{90}{\textbf{Host\textsuperscript{a}}} & \rotatebox{90}{\textbf{Container\textsuperscript{a}}} & \rotatebox{90}{\textbf{Endpoint\textsuperscript{a}}} & \rotatebox{90}{\textbf{Direct\textsuperscript{b}}} & \rotatebox{90}{\textbf{Derived\textsuperscript{b}}} & \rotatebox{90}{\textbf{Point\textsuperscript{c}}} & \rotatebox{90}{\textbf{Cumulative\textsuperscript{c}}} \\
\midrule
cAdvisor & Memory usage & Byte (b) & & \ding{53} & & \ding{53} & & \ding{53} & \\\hdashline
cAdvisor & R/W Disk & Bytes (b) & \ding{53} & & & & \ding{53} & & \ding{53} \\\hdashline
cAdvisor & R/W Network & Bytes (b) & \ding{53} & & & & \ding{53} & & \ding{53} \\\hdashline
Powerjoular & CPU power consumption & Watt (W) & & \ding{53} & & \ding{53} & & \ding{53} & \\\hdashline
Powerjoular & CPU utilization & Percent (\%) & \ding{53} & \ding{53} & & \ding{53} & & \ding{53} & \\\hdashline
perf & DRAM power consumption & Microjoule ($\mu$J) & \ding{53} & & & & \ding{53} & & \ding{53} \\\hdashline
locust & Response time & Millisecond (ms) & & & \ding{53} & \ding{53} & & \ding{53} & \\\hdashline
locust & Failures & Amount & & & \ding{53} & \ding{53} & & \ding{53} & \\
\bottomrule
\multicolumn{10}{l}{\textsuperscript{a}~Scope dimension} \\
\multicolumn{10}{l}{\textsuperscript{b}~Origin dimension} \\
\multicolumn{10}{l}{\textsuperscript{c}~Temporal dimension} \\
\end{tabular}
\end{table}

\section{Experiment Execution} 
\label{sec:execution}

\begin{figure}[htbp]
    \centering
    \includegraphics[width=\columnwidth]{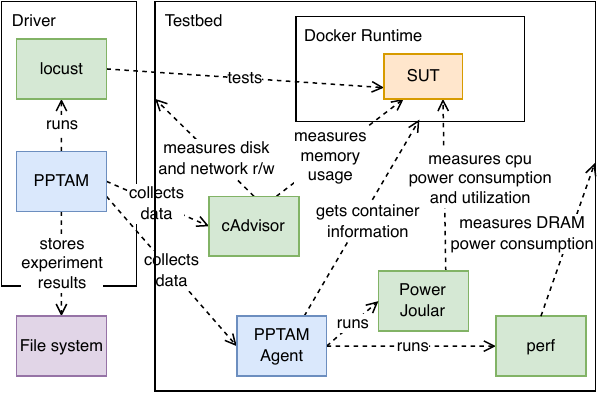}
    \caption{Container diagram of the experimental infrastructure (C4 notation \cite{brown2026}). \cb{0.863}{0.906}{0.980}{0.451}{0.553}{0.733} Blue containers indicate data collection and experiment coordination containers; \cb{0.847}{0.906}{0.835}{0.549}{0.698}{0.431} green containers indicate data collection tools; \cb{0.984}{0.906}{0.812}{0.882}{0.741}{0.404} orange containers indicate the software under test (SUT); and \cb{0.875}{0.835}{0.902}{0.569}{0.455}{0.639} violet containers indicate external systems.}
    \label{fig:architecture}
\end{figure}

The experimental infrastructure consists of two separate machines, as illustrated in Figure \ref{fig:architecture}: a \textit{driver} and a \textit{testbed}. 

The testbed is a machine equipped solely with the software required for data collection and for executing the software under test. Docker containers are constrained to a single virtual CPU (vCPU) and 1 GB of RAM to ensure consistent resource allocation across all experiments. During each experiment, only the application container is running, and Docker services are restarted between consecutive trials to eliminate residual process state and prevent measurement contamination. Concurrent users are ramped up to their target count in under two minutes, establishing a steady load. Following standard practice for performance measurement, data collected during the first two minutes (the warm-up period) are excluded from the analysis to allow the system to reach a stable state.

We operationalize the dependent variables from Sect.~\ref{sec:design} by collecting the metrics listed in Tab.~\ref{tab:metrics}. Response time is measured in milliseconds (ms), CPU usage in percent, memory, disk and network usage in bytes, later converted to mebibytes (MiB) or gibibytes (GiB).

The \textbf{testbed} is a Lenovo X1 Yoga 3rd Gen with an Intel Core i5-8250U CPU and 8~GB of RAM. The software stack consists of Ubuntu 24.04 LTS, PowerJoular~0.4, cAdvisor~0.53, perf~6.8.12, Docker~29.0.2, and the PPTAM \emph{agent} used to collect measurements. The ambient room temperature during the experiments was approx. $25^\circ$C with a variability of $\pm 2^\circ$C.

We measure power consumption using two complementary tools: PowerJoular for per-process CPU power consumption and perf for DRAM energy consumption. Both tools derive their measurements from Intel RAPL counters exposed through the Linux powercap interface \cite{linux_powercap_docs}. Power readings are collected at 1-second intervals throughout each trial, with CPU power and DRAM power expressed in Watts (W).

At runtime, we convert consecutive energy readings from perf into average power by computing the change in energy over the corresponding time interval. This yields the average power between the two readings, expressed in Watts.

CPU utilization is reported as the fraction of the total system CPU capacity. Containers are limited to a single vCPU in Docker. Since the machine has four CPU cores, a value of $0.25$ corresponds to $100\%$ utilization of a single vCPU. The CPU frequency governor is set to \textit{powersave} to reduce variability caused by frequency scaling.

The \textbf{driver} machine is a virtual machine running on Proxmox~VE~8.4.14 with 8~vCPUs and 20~GB of RAM. The software installed on this VM includes Ubuntu~24.04~LTS, PPTAM to orchestrate tests and collect data, and Docker~29.0.2. PPTAM internally uses locust.io~2.42.2. During the experiments, the VM has exclusive access to its allocated resources.

On both machines, unnecessary services and background processes are identified and disabled, and network access is limited to minimize external interference and measurement noise.

\section{Data Analysis}
\label{sec:analysis}


Our analysis proceeds in four steps. First, we conduct a sanity check via trace visualization by plotting the three measurement groups (performance, energy consumption, and resource utilization) for each experiment. This enables us to visually inspect the data, detect recording anomalies, and verify that each antipattern exhibits the characteristic behavior reported in prior work. Second, we perform a descriptive analysis by computing average, minimum, and maximum values of power consumption and response time for each antipattern and generating box plots to obtain an initial view of their distributions. Third, we assess the association between response time and power consumption through Pearson and Spearman correlation coefficients, capturing both linear and monotonic relationships. 

Finally, we fit multiple linear regression models for each experiment, with CPU or DRAM power as the dependent variable and response time, throughput, and CPU utilization as predictors. Using heteroskedasticity-robust (HC3) standard errors, we test the null hypothesis H$_0$ that the response time coefficient is zero; a positive and statistically significant coefficient (p~$< 0.05$) leads us to reject H$_0$, thereby supporting the claim that the corresponding antipattern is associated with increased energy consumption.

\section{Results}
\label{sec:results}

\begin{figure*}[htbp]
    \centering
    \includegraphics[width=.99\linewidth]{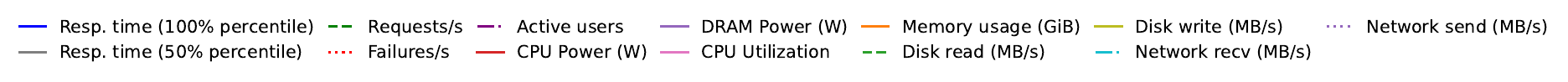}
    \hfill
    \begin{subfigure}{0.34\linewidth}
        \includegraphics[height=161pt]{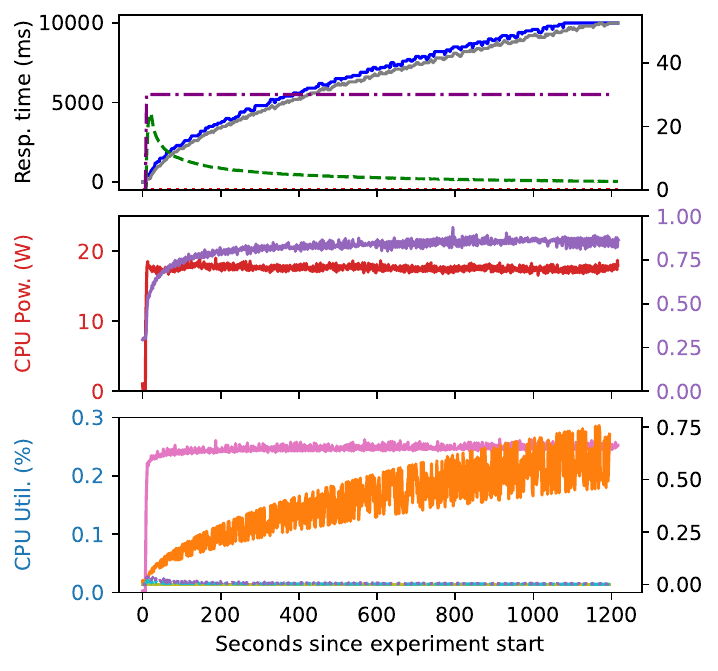}
        \caption{Unnecessary Processing}
        \label{fig:unnecessary_processing}
    \end{subfigure}%
    \begin{subfigure}{0.31\linewidth}
        \centering
        \includegraphics[height=161pt]{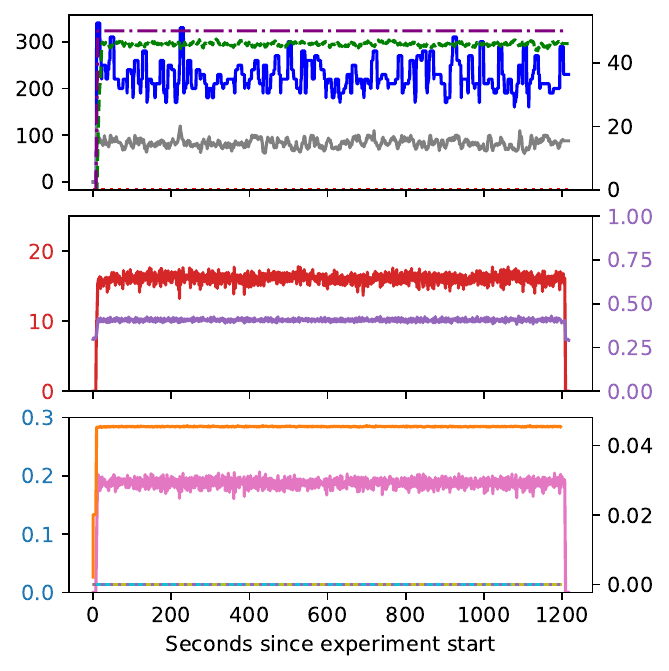}
        \caption{Circuitous Treasure Hunt}
        \label{fig:circuitous_treasure_hunt}
    \end{subfigure}%
    \begin{subfigure}{0.34\linewidth}
        \centering
        \includegraphics[height=161pt]{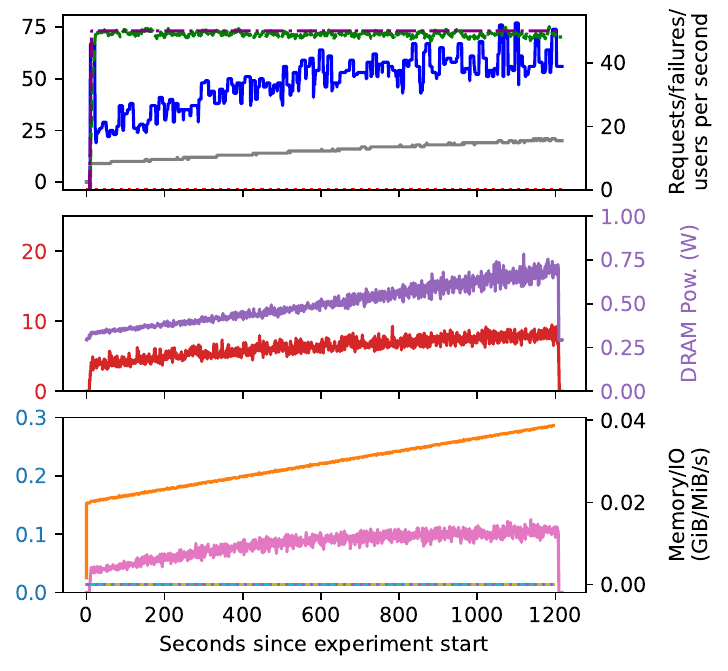}
        \caption{The Ramp}
        \label{fig:the_ramp}
    \end{subfigure}
    \hfill
    \caption{Trace Visualization of Unnecessary Processing, Circuitous Treasure Hunt, and The Ramp.} 
    \label{fig:nice-plots}
\end{figure*}

Beginning with \textit{Sanity Check via Trace Visualization}, in Fig. \ref{fig:nice-plots} we illustrate examples of collected metrics during the respective experiments. The complete set of plots for every antipattern are present in the replication package \cite{replication_package}.

According to our experimental design, the failure rate remains on zero over the entire execution time, and the CPU utilization reaches at least 0.075, which in our testbed corresponds to approximately 25\% of a single vCPU.

In Figure~\ref{fig:unnecessary_processing}, we observe that even with a significant drop in requests per second, there is no corresponding increase in the number of errors. This substantial reduction in throughput therefore has no impact on the error rate, and for this reason the results of this experiment are considered valid.

During the execution of the Circuitous Treasure Hunt antipattern (Figure~\ref{fig:circuitous_treasure_hunt}), the response time fluctuates, while the error rate remains stable at zero. We also observe a small fluctuation in the number of requests per second, which is consistent with the expected behavior of this antipattern.

The Ramp antipattern in Figure~\ref{fig:the_ramp} exhibits the characteristic pattern of steadily increasing metric values, which is precisely how this antipattern is expected to manifest. Nevertheless, in this case as well, both the error rate and CPU utilization satisfy our validity conditions.

These observations collectively support the faithfulness of our Python implementation of the considered antipatterns.


Proceeding to \textit{Descriptive analysis}, Table \ref{tab:descriptive} summarizes the average, minimum, and maximum values of response time, CPU power, and DRAM power consumption across all antipattern experiments. The results reveal substantial variability in response time characteristics. Several antipatterns, such as Unbalanced Processing, God Class, and Excessive Dynamic Allocation, exhibit high average response times (7--14s), while The Ramp and Circuitous Treasure Hunt show response times that remain well below 300 ms. Maximum response time values follow the same pattern, with the most severe antipatterns reaching up to 17s.

\begin{table}[tbhp]
    \centering
    \setlength{\tabcolsep}{2pt}
    \caption{Average, minimum, and maximum values of power consumption and response time for each antipattern. Minimum response time and minimum CPU omitted since they are always 0.}
    \begin{tabular}{@{}L{3.3cm}rrrrrrr@{}}
    \toprule
    & \multicolumn{2}{c}{\textbf{Reponse time}} & \multicolumn{5}{c}{\textbf{Power (W)}} \\
    & \multicolumn{2}{c}{\textbf{(ms)}} & \multicolumn{2}{c}{\textbf{CPU}} & \multicolumn{3}{c}{\textbf{DRAM}} \\
    \textbf{Experiment} & \textbf{$\bar{x}$} & \textbf{$x_{\max}$} & \textbf{$\bar{x}$} & \textbf{$x_{\max}$} & \textbf{$\bar{x}$} & \textbf{$x_{\min}$} & \textbf{$x_{\max}$} \\
    \midrule

Unbalanced Processing & 12833 & 14000 & 18.74 & 23.98 & 0.29 & 0.28 & 0.91 \\\hdashline
Unnecessary Processing & 6728 & 11000 & 17.39 & 22.51 & 0.83 & 0.29 & 1.65 \\\hdashline
The Ramp & 49 & 94 & 6.21 & 12.18 & 0.50 & 0.29 & 1.32 \\\hdashline
Sisyphus DB Retrieval & 2185 & 2500 & 16.81 & 22.40 & 0.76 & 0.29 & 1.07 \\\hdashline
More Is Less & 2289 & 2800 & 18.08 & 23.26 & 0.31 & 0.29 & 0.43 \\\hdashline
God Class & 13693 & 17000 & 13.93 & 22.41 & 0.30 & 0.29 & 0.62 \\\hdashline
Excessive Dyn. Allocation & 7747 & 8200 & 19.36 & 24.74 & 0.29 & 0.28 & 0.69 \\\hdashline
Circuitous Treasure Hunt & 229 & 390 & 15.92 & 22.10 & 0.40 & 0.29 & 1.34 \\\hdashline
One Lane Bridge & 3985 & 4600 & 18.88 & 23.86 & 0.30 & 0.28 & 1.20 \\\hdashline
Traffic Jam & 4139 & 10000 & 17.48 & 23.40 & 0.30 & 0.28 & 1.05 \\

    \bottomrule
    \end{tabular}
    \label{tab:descriptive}
\end{table}

CPU power shows a narrower range across experiments. Mean CPU load typically falls between 14--19W, with peak values consistently reaching 22–25W, indicating that most antipatterns drive the system close to full CPU usage during the workload’s high-load phases.

DRAM power consumption exhibits moderate variability. Mean DRAM power remains low for most experiments (approximately 0.29--0.40W), but some antipatterns such as Unnecessary Processing and Sisyphus DB Retrieval show markedly higher average values (above 0.70W). Maximum DRAM power ranges from 0.43W (More Is Less) up to 1.65W (Unnecessary Processing), suggesting that certain antipatterns cause more pronounced memory-side energy demands.

Overall, the results indicate that while CPU utilization is relatively similar across antipatterns, response time and DRAM power consumption vary considerably, suggesting that different performance antipatterns impose distinct types and intensities of resource stress on the system.

\begin{figure*}[t]
    \centering
    \includegraphics[width=.85\linewidth]{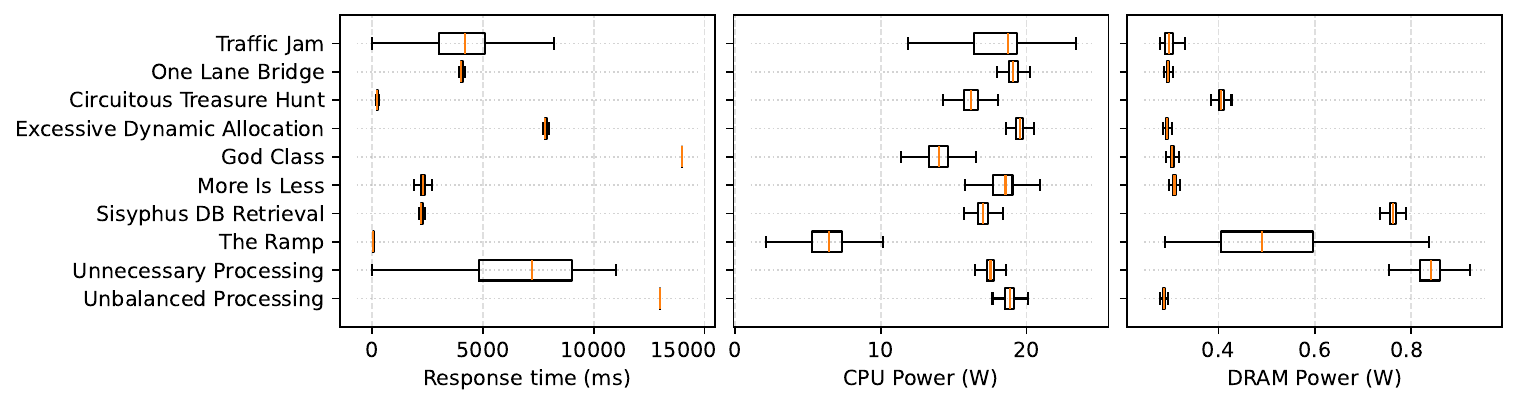}
    \caption{Response time, CPU and DRAM power box plots across all 30 runs for every studied antipattern; boxes show median/IQR, whiskers show central spread, mean marked (outliers hidden for clarity).}
    \label{fig:boxplots1}
\end{figure*}

Fig. \ref{fig:boxplots1} illustrates the box-plots across all experiments for response time, CPU power and DRAM power.

The results of the \textit{Correlation analysis} are reported in Tab. \ref{tab:correlation}. Pearson’s correlation measures the linear association between two variables, indicating how strongly they vary in a proportional, straight-line manner. Spearman’s rank correlation captures monotonic relationships, detecting consistent increasing or decreasing trends even when they are non-linear and being less sensitive to outliers. Using both metrics provides complementary insight: agreement between them suggests a stable relationship, whereas divergence indicates potential non-linearity or the influence of extreme values. We indicate agreement using underlined values. 

\begin{table}[htbp]
    \centering
    \setlength{\tabcolsep}{2pt}
    \caption{Pearson and Spearman correlations between CPU and DRAM power and response time for each experiment (30-run aggregates); Underlined values indicate that both correlations share the same sign.}
    \begin{tabular}{@{}L{2.9cm}rrrr@{}}
    \toprule
    & \multicolumn{2}{c}{\textbf{CPU}} & \multicolumn{2}{c}{\textbf{DRAM}} \\
    \textbf{Experiment} & \textbf{Pearson \(r\)} & \textbf{Spearman $\rho$} & \textbf{Pearson \(r\)} & \textbf{Spearman $\rho$} \\
    \midrule
    Unbalanced Processing & \underline{0.797} & \underline{0.093} & \underline{-0.113} & \underline{-0.072} \\\hdashline
    Unnecessary Processing & 0.204 & -0.101 & \underline{0.352} & \underline{0.786} \\\hdashline
    The Ramp & \underline{0.664} & \underline{0.691} & \underline{0.721} & \underline{0.748} \\\hdashline
    Sisyphus DB Retrieval & 0.613 & -0.264 & 0.641 & -0.264 \\\hdashline
    More Is Less & 0.526 & -0.094 & 0.112 & -0.018 \\\hdashline
    God Class & \underline{0.587} & \underline{0.045} & \underline{0.068} & \underline{0.043} \\\hdashline
    Excessive Dyn. All. & \underline{0.743} & \underline{0.062} & \underline{-0.108} & \underline{-0.028} \\\hdashline
    Circuitous Treasure Hunt & \underline{0.347} & \underline{0.013} & \underline{0.261} & \underline{0.018} \\\hdashline
    One Lane Bridge & \underline{0.648} & \underline{0.029} & \underline{-0.049} & \underline{-0.040} \\\hdashline
    Traffic Jam & \underline{0.296} & \underline{0.226} & \underline{-0.096} & \underline{-0.158} \\
    \bottomrule
    \end{tabular}
    \label{tab:correlation}
\end{table}

Across the experiments, the correlations between response time and power consumption vary substantially, reflecting the heterogeneous behavior of different antipatterns. Several cases (including The Ramp, Unbalanced Processing, and Excessive Dynamic Allocation) show positive CPU–response time correlations, consistent with increased computational effort accompanying higher response times. In contrast, DRAM correlations are more mixed: some antipatterns (e.g., Unnecessary Processing) exhibit strong positive associations, while others show weak or even negative relationships, indicating that memory activity is not uniformly coupled with response time across patterns. Agreement between Pearson and Spearman coefficients (highlighted by underlined values) suggests that, for those experiments, the relationship is both monotonic and directionally stable, whereas divergences point to non-linear or irregular interactions between performance degradation and power usage.


At this point we proceed to \textit{Regression modeling and hypothesis testing}. We estimate the relationship between response time and power consumption using two ordinary least squares (OLS) regression models for each antipattern experiment, aggregating data across 30 repeated runs. OLS quantifies how a dependent variable changes as a linear function of a set of predictors, such that each regression coefficient represents the average marginal effect of its predictor while holding the remaining variables constant. Both models include an intercept and are fitted only on steady-state samples, excluding a 120-second warm-up interval and ignoring negative power readings. This filtering ensures that transient initialization effects and instrumentation artifacts do not bias the estimated relationship between response time and power consumption, allowing the models to capture behavior that is representative of the system's stable operating mode.

Because system-level performance measurements are typically noisy and exhibit load-dependent variability (see e.g., \cite{Schad2010}), the assumption of homoskedastic residuals (i.e., constant variance in the regression errors) is unlikely to hold. In line with this expectation, the Anderson-Darling test \cite{Anderson01121954} rejected normality for all experiments; the Breusch-Pagan test \cite{Breusch1979} rejected the null hypothesis of homoskedasticity for every antipattern in at least one of the two regression models (CPU or DRAM power). Specifically, all antipatterns except \emph{Unnecessary Processing} showed heteroskedasticity in the CPU--power model, while for DRAM power the null was rejected for all but \emph{More Is Less}, \emph{God Class}, \emph{Excessive Dynamic Allocation}, and \emph{One Lane Bridge}. Thus, each antipattern exhibited evidence of heteroskedasticity in one or the other analysis.

As a consequence, we employ HC3 heteroskedasticity-consistent (robust) standard errors \cite{MACKINNON1985305}. HC3 adjusts the estimated uncertainty of regression coefficients in the presence of outliers or non-uniform noise, yielding more reliable inference than classical OLS standard errors when homoskedasticity does not hold. Using HC3 enables valid hypothesis testing even when measurement variability increases over time. 

For the CPU power model, we regress CPU power on response time, request rate, and CPU utilization. These predictors are chosen to isolate the contribution of response time to CPU energy usage while accounting for factors that directly influence CPU power. Request rate captures the volume of work offered to the system, and CPU utilization reflects the degree of processor saturation; both variables typically rise and fall with workload intensity and may correlate with response time. By including them, the model controls for load-induced variation in CPU power, ensuring that the estimated response time coefficient represents the effect of performance degradation rather than changes in system demand or CPU occupancy. The goal is thus to obtain an estimate of whether higher response time is associated with increased CPU power consumption under otherwise comparable conditions.

For the DRAM power model, we regress DRAM power on response time, request rate, CPU utilization, and memory usage. The same workload controls (request rate and CPU utilization) are included to account for load-induced changes in memory traffic, while memory usage is added because it directly determines the volume of active DRAM state and is a primary driver of DRAM power draw. Including these predictors addresses potential confounding: DRAM activity may rise due to increased memory pressure or higher throughput, independently of response time. By selecting variables with clear mechanistic relevance to DRAM behavior, the model remains parsimonious while enabling a more accurate assessment of whether increased response time, after controlling for workload and memory demand, is associated with higher memory-side energy consumption.

For each experiment (with sample sizes $N > 32{,}000$ in all cases), we report the estimated response time coefficient ($\beta$) together with its 95\% confidence interval and $p$-value (computed using HC3 standard errors), along with a decision on whether the null hypothesis $H_0$, i.e., that no significant relationship exists between system performance and power usage, is rejected or retained. When $H_0$ is rejected, an arrow indicates whether response time has a statistically significant positive or negative effect on power consumption at $\alpha = 0.05$.

The goal of this analysis is to assess whether \textit{increased} response time, a manifestation of performance degradation, corresponds to \textit{higher power} drain once workload intensity and resource utilization are controlled for. A significant positive response time coefficient would indicate that the antipattern is associated with increased power consumption, whereas a nonsignificant or negative coefficient suggests that degraded performance does not necessarily translate into higher energy demand under the tested conditions. The detailed results are shown in Tab.~\ref{tab:cpu-dram-power-rt}.

\begin{table*}[htbp]
    \centering
    \setlength{\tabcolsep}{3pt}
    \caption{CPU and DRAM Power vs Response Time}
    \begin{tabular}{@{}L{3.2cm}rrrrlrrrrl@{}}
    \toprule
    & \multicolumn{5}{c}{\textbf{CPU Power}} 
    & \multicolumn{5}{c}{\textbf{DRAM Power}} \\
    \cmidrule(lr){2-6} \cmidrule(lr){7-11}
    \textbf{Experiment} 
      & \textbf{$\beta_{lat}$} & \textbf{CI low} & \textbf{CI high} & \textbf{$p_{lat}$} & \textbf{Decision}
      & \textbf{$\beta_{lat}$} & \textbf{CI low} & \textbf{CI high} & \textbf{$p_{lat}$} & \textbf{Decision} \\
    \midrule

    Unbalanced Processing & -0.000083 & -0.000147 & -0.000018 & 0.011860 & Reject ↓ &  0.000000 & -0.000000 &  0.000001 & 0.799513 & Keep \\ \hdashline
    Unnecessary Processing & -0.000002 & -0.000008 &  0.000004 & 0.422677 & Keep & -0.000004 & -0.000006 & -0.000002 & 0.000127 & Reject ↓ \\ \hdashline
    The Ramp &  0.018475 &  0.017443 &  0.019508 & 0.000000 & Reject ↑ & -0.000677 & -0.000743 & -0.000610 & 0.000000 & Reject ↓ \\ \hdashline
    Sisyphus DB Retrieval & -0.000115 & -0.000197 & -0.000032 & 0.006348 & Reject ↓ & -0.000005 & -0.000007 & -0.000002 & 0.000231 & Reject ↓ \\ \hdashline
    More Is Less & -0.001140 & -0.001290 & -0.000991 & 0.000000 & Reject ↓ & -0.000001 & -0.000002 & -0.000001 & 0.000007 & Reject ↓ \\ \hdashline
    God Class &  0.000190 &  0.000153 &  0.000227 & 0.000000 & Reject ↑ & -0.000000 & -0.000000 &  0.000000 & 0.141974 & Keep \\ \hdashline
    Excessive Dyn.\ All. &  0.000017 & -0.000026 &  0.000059 & 0.445093 & Keep & -0.000000 & -0.000001 &  0.000001 & 0.964429 & Keep \\ \hdashline
    Circuitous Treasure Hunt & -0.000566 & -0.000771 & -0.000360 & 0.000000 & Reject ↓ &  0.000006 & -0.000001 &  0.000012 & 0.108985 & Keep \\ \hdashline
    One Lane Bridge & -0.000047 & -0.000119 &  0.000024 & 0.194439 & Keep &  0.000000 & -0.000001 &  0.000002 & 0.465744 & Keep \\ \hdashline
    Traffic Jam &  0.000029 &  0.000023 &  0.000035 & 0.000000 & Reject ↑ & -0.000000 & -0.000000 &  0.000000 & 0.814016 & Keep \\

    \bottomrule
    \end{tabular}
    \label{tab:cpu-dram-power-rt}
\end{table*}
%
%

\section{Discussion}


\textbf{Our study demonstrates that when certain performance antipatterns are present, we can indeed observe a change in the the power consumption; depending on the pattern, we observe either an increase or decrease.}
We observe that power consumption growth is primarily influenced by workload intensity rather than by the accumulated execution time alone.
For instance, the Circuitous Treasure Hunt antipattern, illustrated in Figure \ref{fig:circuitous_treasure_hunt}, increases CPU usage around 80\% quickly, capping power draw at an upper bound, even as response time holds steady in a range below 300 ms. 
This indicates energy use ties to intensity, not accumulated response time, \textit{so no additional waste accrues from time alone}, though the antipattern itself represents inherent inefficiency.
This observation is confirmed in Table \ref{tab:cpu-dram-power-rt}, where the correlation analysis shows the limited effect of response time on CPU Power.
However, we expect that refactoring the antipattern could lower the observed power/CPU usage below saturation.
The scenario provided by the Circuitous Treasure Hunt represents a relatively favorable case from an energy perspective, because energy is not additionally wasted through extra execution time caused by progressive slowdowns.
Despite a CPU usage spike of around 80\%, this antipattern is not inherently CPU-intensive.
The observed response time is therefore likely due to its implementation.
Specifically, our implementation performs multiple sequential database calls, which primarily stress disk and memory resources rather than the CPU.
As a result, the way data is retrieved and joined across different tables has a major impact on response time.
Optimizing query batching or introducing asynchronous requests could further reduce unnecessary CPU activity while maintaining the intended functionality and response time characteristics of the scenario.

This behavior differs from that of Unnecessary Processing, shown in Figure \ref{fig:unnecessary_processing}. 
Like Circuitous Treasure Hunt, it saturates CPU usage and limits power consumption.
However, response time increases steadily throughout the observation period.
As a result, although the software system may appear energy-efficient under full CPU utilization, it exhibits poor responsiveness as response time continues to grow.
Unlike Circuitous Treasure Hunt, this antipattern performs CPU-intensive operations, repeatedly carrying out redundant hash computations and input validations. 
It also continuously allocates and reorders data in memory, resulting in a sequence of superfluous operations whose footprint gradually expands over time. 
This ongoing growth in unnecessary processing causes increasing response time, resembling a mild instance of The Ramp antipattern.
This observation suggests that, depending on the implementation, \textit{multiple antipatterns may manifest simultaneously}.

Our implementation of The Ramp also comes with additional insights for practitioners.
In Figure \ref{fig:the_ramp} we observe a typical Ramp situation: both CPU and DRAM power consumption increase steadily over time, and this trend is mirrored by the growth in response time.
By definition, The Ramp antipattern is characterized by a workload that becomes progressively heavier during the observation window. In our case, as more data are accumulated in memory, each new request must process a larger in-memory dataset.
In the concrete implementation, this behaviour is caused by continuously storing additional data structures and then performing search and ordering operations on them.
As the in-memory collections grow, these operations become increasingly expensive, both in terms of computation and memory access.
Consequently, the application performs more work per request, leading to longer response times and a gradual rise in power consumption for both CPU and DRAM, fully aligned with the expected profile of The Ramp antipattern.
The pattern implies long-term scalability and capacity issues: since each request becomes more expensive as the system accumulates data, the system will not only slow down but also show decreasing throughput over time, which is a defining characteristic of The Ramp antipattern. 
This means that simply adding hardware will not solve the problem sustainably, because the fundamental per-request work keeps increasing with dataset size.
\textit{This indicates that refactoring data structures or choosing algorithms whose cost grows more slowly with dataset size (e.g., indexed access, partitioning, batching) is at least as important as restructuring workload over time}.

We suggest to practitioners that optimization strategies should consider \textbf{trade-offs} between intensity (e.g., power and CPU usage) metrics and response time to detect hidden production costs in the future.
By analysing Figure~\ref{fig:boxplots1}, we observe that CPU power consumption is relatively stable across all antipatterns except for two, while the response time depends on the different cases.
The Ramp shows clear variability within a low power-consumption range. This behavior is expected, as the system is not in a critical condition and the antipattern still allows it to operate in a near-normal regime.
This interpretation is further supported by the response time, which remains remarkably stable.
Traffic Jam exhibits CPU power values ranging from medium to near-maximum levels.
A similar variability is visible in the response time distribution. This behaviour reflects alternating phases in which incoming requests are easily handled and phases during the jam in which the system is over-stressed and must process accumulated backlogs.
The scenario presented by Circuitous Treasure Hunt is particularly challenging to optimize, as the system seems already to be operation at a condition to obtain the best response time/resource usage trade-off.
Since the system already operates at saturation, increasing workload intensity (i.e., adding more concurrent users) is likely to raise response time and degrade performance.
Conversely, restructuring the workload (i.e., adjusting the user arrival rate) may not enhance energy efficiency, as the application is already using its available resources. 

We observe that different antipatterns impose varying energy loads on the system.
Although energy consumption (energy refers to the total amount of work done over time, whereas power is the instantaneous rate of consumption) is not the primary focus of this study, it is affected by numerous confounding factors, such as the number of users, the underlying hardware platform, and the specific implementation details of each antipattern.
This work seeks to examine the relationship between power consumption, response time, and resource utilization.
Notably, our findings indicate that antipatterns executed under similar conditions exhibit distinct energy loads, \textit{offering a promising foundation for further investigation into the energy implications of performance antipatterns}. 
Table \ref{tab:energy} presents the average energy consumption associated with each antipattern, computed using the trapezoidal rule using the power measurements.  
We notice that antipatterns executed with the same number of concurrent users (50), that is, all except Unbalanced Processing (10), Unnecessary Processing (30), and Traffic Jam (30), demonstrate different CPU and DRAM consumption.
Excessive Dynamic Allocation results in the highest CPU energy usage, while Unnecessary Processing yields the greatest DRAM energy load.
These findings are valuable for practitioners evaluating whether mitigating a given antipattern can help achieve a more balanced trade-off between energy efficiency and performance.
For instance, God Class and The Ramp introduce significantly lower energy overhead compared to the other antipatterns.


\begin{table}[htbp]
    \centering
    \setlength{\tabcolsep}{2pt}
    \caption{Average CPU and DRAM energy consumption.}
    \label{tab:energy}

    \begin{tabular}{@{}lcc@{}}
    \toprule
    \textbf{Experiment} & \textbf{CPU (kJ)} & \textbf{DRAM (kJ)} \\
    \midrule
    Unbalanced Processing        & 21.933541 & 0.315958 \\\hdashline
    Unnecessary Processing       & 20.528362 & 0.909147 \\\hdashline
    The Ramp                     & 10.620640 & 0.569759 \\\hdashline
    Sisyphus DB Retrieval        & 19.959613 & 0.834051 \\\hdashline
    More Is Less                 & 20.986992 & 0.339830 \\\hdashline
    God Class                    & 17.951567 & 0.335021 \\\hdashline
    Excessive Dynamic Allocation & 22.727046 & 0.322975 \\\hdashline
    Circuitous Treasure Hunt     & 19.252548 & 0.445187 \\\hdashline
    One Lane Bridge              & 22.227476 & 0.325174 \\\hdashline
    Traffic Jam                  & 20.834088 & 0.327770 \\
    \bottomrule
    \end{tabular}
\end{table}

\section{Threats To Validity}

\textit{Internal Validity:} Each performance antipattern is manually implemented and may introduce unintended side effects. To reduce this risk, we apply the tactic to visually validate the behavior of every antipattern against its description in \cite{smith2003more} using line plots of response time, CPU utilization, and power consumption (see Fig.~\ref{fig:the_ramp}). Between each experiment, applications containing the antipattern are uninstalled, the entire Docker environment restarted, the applications re-installed.

\textit{External Validity:} Our results are obtained from Python-based containers running on a single hardware configuration. This controlled setup improves reproducibility but limits generalizability to other platforms, languages, and deployment environments. Real-world systems often exhibit more heterogeneous workloads and resource variability, which may influence the observed effects. Our tactic to reduce this risk is to provide the implemented antipatterns together with all collected data in form of a replication package \cite{replication_package}.

\textit{Construct Validity:} Because \cite{smith2003more} describes performance antipatterns only conceptually, their exact implementations are not defined. The tactic to mitigate this risk is by confirming that each microservice exhibits the expected resource-usage and response time patterns as well as providing the rationale for their implementation in Tab. \ref{tab:implementations_rationale}.

\textit{Conclusion Validity:} Measurement accuracy and statistical inference represent additional threats. DRAM power readings, collected via RAPL, reflect system-wide consumption and cannot be isolated per container, though we eliminate competing workloads to limit noise. Each experiment is repeated 30 times, warm-up periods are excluded, and HC3 robust standard errors are used to account for non-normality and heteroskedasticity in the residuals. Furthermore, Table~\ref{tab:implementations_rationale} documents the concrete implementation choices and their rationale for each antipattern, enabling verification, replication, and adaptation of our implementations by other researchers.

\section{Conclusion and Future Work}
Software performance antipatterns are recurring situations in software design that are known to degrade system performance.
Although these antipatterns are well recognized, there is limited evidence regarding their impact on software energy efficiency.
This study investigates how software performance relates to energy efficiency by analyzing the relationship between response time and power consumption.
We conducted a controlled experiment using the ten performance antipatterns presented by~\citet{Smith2002}.
Each antipattern is implemented as a containerized service, subjected to a fixed workload, and profiled to measure performance and power metrics.
The results show that performance antipatterns do not have a uniform effect on energy consumption.
In some cases, performance issues that increase response time also lead to greater energy waste, such as in The Ramp antipattern.
In contrast, other antipatterns, such as Unnecessary Processing, do not significantly affect response time, which remains within a constant range, despite higher CPU utilization and power draw.
We encourage practitioners to consider response time and energy consumption as independent but interrelated metrics when evaluating software performance.
As future work, we plan to inject these antipatterns into established microservice benchmarks (e.g., Train Ticket, TeaStore) and replicate the analysis at the system level and across different platforms.
This will enable the identification of design patterns that promote better trade-offs between performance and energy efficiency.

\balance

\newpage
\bibliographystyle{IEEEtranN}
\bibliography{reference}

\end{document}